\documentstyle[prd,aps,preprint,tighten,epsfig]{revtex}
\begin{document}
\draft

\title{\Large\bf Rephasing Invariants of $CP$ and $T$ Violation \\
in the Four-Neutrino Mixing Models}
\author{{\bf Wan-lei Guo} ~ and ~ {\bf Zhi-zhong Xing}}
\address{Institute of High Energy Physics, P.O. Box 918 (4),
Beijing 100039, China \\
({\it Electronic address: xingzz@mail.ihep.ac.cn})}
\maketitle

\begin{abstract}
We calculate the rephasing invariants of $CP$ and $T$ violation
in a favorable parametrization of the $4\times 4$
lepton flavor mixing matrix. Their relations with the
$CP$- and $T$-violating asymmetries in neutrino oscillations are
derived, and the matter effects are briefly discussed.
\end{abstract}



Since 1998, some robust evidence for atmospheric and solar neutrino
oscillations has been accumulated from the Super-Kamiokande
experiment \cite{SK98}. In addition, $\nu_\mu \rightarrow \nu_e$
and $\overline{\nu}_\mu \rightarrow \overline{\nu}_e$ transitions
have been observed by the LSND Collaboration \cite{LSND}.
A simultaneous interpretation of solar, atmospheric and LSND data
requires the introduction of a light sterile neutrino \cite{4N}, because
they involve three distinct mass-squared differences:
$\Delta m^2_{\rm LSND} \gg \Delta m^2_{\rm atm} \gg \Delta m^2_{\rm sun}$.
In the four-neutrino mixing models, $CP$ and $T$ symmetries are generally
expected to be violated. The measurement of leptonic $CP$- and $T$-violating
effects needs a new generation of accelerator neutrino experiments with
very long baselines. So far some attention has been paid to the
possibilities to observe four-neutrino mixing and $CP$ (or $T$) violation
in a variety of long-baseline neutrino oscillation
experiments \cite{Dai,Tanimoto,Gavela,XingSUM,Hattori}.

The violation of $CP$ and $T$ invariance in neutrino oscillations is
attributed to the nontrivial complex phases in the $4\times 4$ lepton
flavor mixing matrix $V$. While the $CP$-violating phases of $V$ can be
assigned in many different ways, the observable effects of $CP$ and $T$
violation depend only upon some rephasing invariants of $V$ \cite{Jarlskog}.
It is therefore useful to investigate
how those rephasing invariants are related to the flavor mixing
angles and $CP$-violating phases in a specific parametrization of $V$,
and how they are connected to the $CP$ and $T$ asymmetries in
long-baseline neutrino oscillations. Although some attempts were
made for this purpose \cite{Dai,Tanimoto,Gavela,XingSUM},
a complete and analytically exact result has not been achieved.

In this paper we establish the relationship between the
$CP$ and $T$ asymmetries in neutrino oscillations and the rephasing
invariants of $CP$ and $T$ violation defined in the four-neutrino mixing
models. In particular, the exact expressions of those rephasing-invariants
are for the first time calculated on the basis of a favorable
parametrization of the $4\times 4$ lepton flavor mixing matrix $V$.
Our results are very useful for a systematic and model-independent
study of $CP$- and $T$-violating effects in various long-baseline neutrino
oscillation experiments, and some of them are also applicable for the
four-quark mixing models.

Let us begin with a generic $\rm SU(2)_L \times U(1)_Y$ model of electroweak
interactions, in which there exist $n$ charged leptons belonging to
isodoublets, $n$ active neutrinos belonging to isodoublets, and $n'$
sterile neutrinos belonging to isosinglets. The charged-current weak
interactions of leptons are then associated with a rectangular flavor
mixing matrix of $n$ rows and $(n+n')$ columns \cite{SV}.
Without loss of generality, one may choose to identify the flavor
eigenstates of charged leptons with their mass eigenstates. In this
specific basis, the $n\times (n+n')$ lepton mixing matrix links the neutrino
flavor eigenstates directly to the neutrino mass eigenstates. Although
sterile neutrinos do not participate in normal weak interactions, they
may oscillate among themselves and with active neutrinos. Once the latter
is concerned we are led to a more general $(n+n')\times (n+n')$ lepton
flavor mixing matrix \cite{FXreview}, defined as $V$ in the chosen
flavor basis. For the flavor mixing
of one sterile neutrino ($\nu_s$) and three active neutrinos
($\nu_e, \nu_\mu, \nu_\tau$), the explicit form of $V$ can be written as
\begin{equation}
\left ( \matrix{
\nu_s \cr
\nu_e \cr
\nu_\mu \cr
\nu_\tau \cr} \right ) \; = \; \left ( \matrix{
V_{s0} & V_{s1} & V_{s2} & V_{s3} \cr
V_{e0} & V_{e1} & V_{e2} & V_{e3} \cr
V_{\mu 0} & V_{\mu 1} & V_{\mu 2} & V_{\mu 3} \cr
V_{\tau 0} & V_{\tau 1} & V_{\tau 2} & V_{\tau 3} \cr} \right )
\left ( \matrix{
\nu_0 \cr
\nu_1 \cr
\nu_2 \cr
\nu_3 \cr} \right ) \; ,
\end{equation}
where $\nu_i$ (for $i=0,1,2,3$) denote the mass eigenstates of four neutrinos.
If neutrinos are Dirac particles, $V$ can be parametrized in terms of six
mixing angles and three phase angles. If neutrinos are Majorana particles,
however, three additional phase angles are needed to get a full
parametrization of $V$. In either case, one may define
the rephasing invariants of $CP$ or $T$ violation as follows:
\begin{equation}
J^{ij}_{\alpha\beta} \; \equiv \; {\rm Im} \left ( V_{\alpha i} V_{\beta j}
V^*_{\alpha j} V^*_{\beta i} \right ) \; ,
\end{equation}
where the Greek subscripts run over $(s, e, \mu, \tau)$ and the Latin
superscripts run over $(0, 1, 2, 3)$. Of course,
$J^{ii}_{\alpha\beta} = J^{ij}_{\alpha\alpha} =0$ and
$J^{ij}_{\alpha\beta} = -J^{ji}_{\alpha\beta}
= -J^{ij}_{\beta\alpha} = J^{ji}_{\beta\alpha}$ hold by definition.
The unitarity of $V$ leads to the following correlation equations of
$J^{ij}_{\alpha \beta}$:
\begin{equation}
\sum_i J^{ij}_{\alpha\beta} \; = \;
\sum_j J^{ij}_{\alpha\beta} \; = \;
\sum_\alpha J^{ij}_{\alpha\beta} \; = \;
\sum_\beta J^{ij}_{\alpha\beta} \; =\; 0 \; .
\end{equation}
Hence there are totally nine independent $J^{ij}_{\alpha\beta}$, whose
magnitudes depend only upon three of the six $CP$-violating phases or
their combinations in a specific parametrization of $V$.
Let us follow Ref. \cite{Dai} to parametrize $V$ as
\small
\begin{equation}
V \; = \; \left ( \matrix{
c_{01}c_{02}c_{03}
& c_{02}c_{03}\hat{s}_{01}^*
& c_{03}\hat{s}_{02}^*
& \hat{s}_{03}^*
\cr\cr
-c_{01}c_{02}\hat{s}_{03}\hat{s}_{13}^*
& -c_{02}\hat{s}_{01}^*\hat{s}_{03}\hat{s}_{13}^*
& -\hat{s}_{02}^*\hat{s}_{03}\hat{s}_{13}^*
& c_{03}\hat{s}_{13}^*
\cr
-c_{01}c_{13}\hat{s}_{02}\hat{s}_{12}^*
& -c_{13}\hat{s}_{01}^*\hat{s}_{02}\hat{s}_{12}^*
& +c_{02}c_{13}\hat{s}_{12}^*
&
\cr
-c_{12}c_{13}\hat{s}_{01}
& +c_{01}c_{12}c_{13}
&
&
\cr\cr
-c_{01}c_{02}c_{13}\hat{s}_{03}\hat{s}_{23}^*
& -c_{02}c_{13}\hat{s}_{01}^*\hat{s}_{03}\hat{s}_{23}^*
& -c_{13}\hat{s}_{02}^*\hat{s}_{03}\hat{s}_{23}^*
& c_{03}c_{13}\hat{s}_{23}^*
\cr
+c_{01}\hat{s}_{02}\hat{s}_{12}^*\hat{s}_{13}\hat{s}_{23}^*
& +\hat{s}_{01}^*\hat{s}_{02}\hat{s}_{12}^*\hat{s}_{13}\hat{s}_{23}^*
& -c_{02}\hat{s}_{12}^*\hat{s}_{13}\hat{s}_{23}^*
&
\cr
-c_{01}c_{12}c_{23}\hat{s}_{02}
& -c_{12}c_{23}\hat{s}_{01}^*\hat{s}_{02}
& +c_{02}c_{12}c_{23}
&
\cr
+c_{12}\hat{s}_{01}\hat{s}_{13}\hat{s}_{23}^*
& -c_{01}c_{12}\hat{s}_{13}\hat{s}_{23}^*
&
&
\cr
+c_{23}\hat{s}_{01}\hat{s}_{12}
& -c_{01}c_{23}\hat{s}_{12}
&
&
\cr\cr
-c_{01}c_{02}c_{13}c_{23}\hat{s}_{03}
& -c_{02}c_{13}c_{23}\hat{s}_{01}^*\hat{s}_{03}
& -c_{13}c_{23}\hat{s}_{02}^*\hat{s}_{03}
& c_{03}c_{13}c_{23}
\cr
+c_{01}c_{23}\hat{s}_{02}\hat{s}_{12}^*\hat{s}_{13}
& +c_{23}\hat{s}_{01}^*\hat{s}_{02}\hat{s}_{12}^*\hat{s}_{13}
& -c_{02}c_{23}\hat{s}_{12}^*\hat{s}_{13}
&
\cr
+c_{01}c_{12}\hat{s}_{02}\hat{s}_{23}
& +c_{12}\hat{s}_{01}^*\hat{s}_{02}\hat{s}_{23}
& -c_{02}c_{12}\hat{s}_{23}
&
\cr
+c_{12}c_{23}\hat{s}_{01}\hat{s}_{13}
& -c_{01}c_{12}c_{23}\hat{s}_{13}
&
&
\cr
-\hat{s}_{01}\hat{s}_{12}\hat{s}_{23}
& +c_{01}\hat{s}_{12}\hat{s}_{23}
&
&
\cr } \right ) \; ,
\end{equation}
\normalsize
where $c_{ij} \equiv \cos \theta_{ij}$ and
$\hat{s}_{ij} \equiv s_{ij} e^{{\rm i}\delta_{ij}}$ with
$s_{ij} \equiv \sin \theta_{ij}$. Without loss of generality, the
six mixing angles $\theta_{ij}$ can all be arranged to lie in the
first quadrant. The six $CP$-violating phases $\delta_{ij}$ may
take arbitrary values between $0$ and $2\pi$.
After some lengthy but straightforward
calculations, we obtain the explicit expressions of nine independent
$J^{ij}_{\alpha \beta}$ defined in Eq. (2):
\small
\begin{eqnarray}
J^{02}_{\tau s} & = & c^2_{01} c_{02} c^2_{03} c_{12} c_{13} c_{23}
s_{02} s_{03} s_{23} \sin \phi_x + c_{01} c_{02} c^2_{03} c_{12} c_{13}
c^2_{23} s_{01} s^2_{02} s_{03} s_{13} \sin \phi_y
\nonumber \\
&& + \left (c^2_{23} s^2_{13} - s^2_{23} \right ) c_{01} c^2_{02} c^2_{03}
c_{12} s_{01} s_{02} s_{12} \sin \phi_z - c_{01} c^2_{02} c^2_{03}
c^2_{12} c_{23} s_{01} s_{02} s_{13} s_{23} \sin (\phi_x - \phi_y)
\nonumber \\
&& - c_{01} c_{02} c^2_{03} c_{13} c_{23} s_{01} s^2_{02} s_{03}
s_{12} s_{23} \sin (\phi_x + \phi_z) + c^2_{01} c_{02} c^2_{03} c_{13}
c^2_{23} s_{02} s_{03} s_{12} s_{13} \sin (\phi_y - \phi_z)
\nonumber \\
&& - c_{01} c^2_{02} c^2_{03} c_{23} s_{01} s_{02} s^2_{12}
s_{13} s_{23} \sin (\phi_x - \phi_y + 2\phi_z) \; ,
\nonumber \\
J^{03}_{\tau s} & = & -c^2_{01} c_{02} c^2_{03} c_{12} c_{13}
c_{23} s_{02} s_{03} s_{23} \sin \phi_x -
c_{01} c_{02} c^2_{03} c_{12} c_{13} c^2_{23} s_{01} s_{03} s_{13}
\sin \phi_y
\nonumber \\
&& + c_{01} c_{02} c^2_{03} c_{13} c_{23} s_{01} s_{03} s_{12}
s_{23} \sin (\phi_x + \phi_z) - c^2_{01} c_{02} c^2_{03} c_{13} c^2_{23}
s_{02} s_{03} s_{12} s_{13} \sin (\phi_y - \phi_z) \; ,
\nonumber \\
J^{23}_{\tau s} & = & c_{02} c^2_{03} c_{12} c_{13} c_{23} s_{02}
s_{03} s_{23} \sin \phi_x + c_{02} c^2_{03} c_{13} c^2_{23}
s_{02} s_{03} s_{12} s_{13} \sin (\phi_y - \phi_z) \; ;
\end{eqnarray}
\normalsize
and
\small
\begin{eqnarray}
J^{02}_{se} & = & c_{01} c_{02} c^2_{03} c_{12} c_{13} s_{01}
s^2_{02} s_{03} s_{13} \sin \phi_y - c_{01} c^2_{02} c^2_{03}
c_{12} c^2_{13} s_{01} s_{02} s_{12} \sin \phi_z
\nonumber \\
&& + c^2_{01} c_{02} c^2_{03} c_{13} s_{02} s_{03} s_{12} s_{13}
\sin (\phi_y - \phi_z) \; ,
\nonumber \\
J^{13}_{se} & = & c_{01} c_{02} c^2_{03} c_{12} c_{13} s_{01} s_{03}
s_{13} \sin \phi_y - c_{02} c^2_{03} c_{13} s^2_{01} s_{02} s_{03}
s_{12} s_{13} \sin (\phi_y - \phi_z) \; ,
\nonumber \\
J^{23}_{se} & = & c_{02} c^2_{03} c_{13} s_{02} s_{03} s_{12} s_{13}
\sin (\phi_y - \phi_z) \; ;
\end{eqnarray}
\normalsize
as well as
\small
\begin{eqnarray}
J^{12}_{e\mu} & = & - \left ( c^2_{01} c^2_{12} s^2_{13} -
c^2_{01} c^2_{13} s^2_{12} - s^2_{01} s^2_{03} s^2_{13} +
s^2_{01} s^2_{12} \right ) c_{02} c_{12} c_{13} c_{23} s_{02}
s_{03} s_{23} \sin \phi_x
\nonumber \\
&& + \left ( c^2_{02} c^2_{23} s^2_{12} - c^2_{12} c^2_{23} s^2_{02}
+ s^2_{02} s^2_{03} s^2_{23} - s^2_{12} s^2_{23} \right )
c_{01} c_{02} c_{12} c_{13} s_{01} s_{03} s_{13} \sin \phi_y
\nonumber \\
&& + \left ( c^2_{13} c^2_{23} - c^2_{13} s^2_{03} s^2_{23}
- c^2_{23} s^2_{03} s^2_{13} + s^2_{03} s^2_{13} s^2_{23}  \right )
c_{01} c^2_{02} c_{12} s_{01} s_{02} s_{12} \sin \phi_z
\nonumber \\
&& + \left ( c^2_{02} s^2_{13} - c^2_{13} s^2_{02} \right )
c_{01} c^2_{12} c_{23} s_{01} s_{02} s^2_{03} s_{13} s_{23}
\sin (\phi_x - \phi_y)
\nonumber \\
&& + \left ( c^2_{12} - c^2_{02} s^2_{13} - c^2_{13} s^2_{02} +
s^2_{02} s^2_{03} s^2_{13} \right ) c_{01} c_{02} c_{13} c_{23} s_{01} s_{03}
s_{12} s_{23} \sin (\phi_x + \phi_z)
\nonumber \\
&& - \left ( c^2_{01} c^2_{12} c^2_{23} - c^2_{01} c^2_{12} s^2_{23}
- c^2_{12} c^2_{23} s^2_{01} + s^2_{01} s^2_{03} s^2_{23}
- s^2_{01} s^2_{12} s^2_{23} \right ) c_{02} c_{13} s_{02} s_{03}
s_{12} s_{13} \sin (\phi_y - \phi_z)
\nonumber \\
&& + \left ( c^2_{01} c^2_{02} c^2_{13} - c^2_{01} c^2_{13} s^2_{02} s^2_{03}
- c^2_{02} c^2_{13} s^2_{01} s^2_{03} + c^2_{13} s^2_{01} s^2_{02}
s^2_{03} \right ) c_{12} c_{23} s_{12} s_{13} s_{23}
\sin (\phi_x - \phi_y + \phi_z)
\nonumber \\
&& - \left ( c^2_{02} c^2_{13} s^2_{12} -
c^2_{02} s^2_{03} s^2_{12} s^2_{13} -
c^2_{13} s^2_{02} s^2_{03} s^2_{12} \right )
c_{01} c_{23} s_{01} s_{02} s_{13} s_{23}
\sin (\phi_x - \phi_y + 2\phi_z)
\nonumber \\
&& - c_{01} c^2_{02} c^2_{13} c_{23} s_{01} s_{02} s^2_{03} s_{13} s_{23}
\sin (\phi_x + \phi_y)
+ c_{01} c_{02} c^2_{12} c_{13} c_{23} s_{01} s^2_{02} s_{03}
s_{12} s_{23} \sin (\phi_x - \phi_z)
\nonumber \\
&& + \left ( c^2_{23} - s^2_{23} \right ) c_{01} c_{02} c_{12} c_{13}
s_{01} s^2_{02} s_{03} s^2_{12} s_{13} \sin (\phi_y - 2\phi_z)
\nonumber \\
&& - \left (c^2_{01} - s^2_{01} \right ) c_{02} c_{12} c_{13} c_{23}
s_{02} s_{03} s^2_{12} s^2_{13} s_{23}
\sin (\phi_x - 2\phi_y + 2\phi_z)
\nonumber \\
&& - c_{01} c_{02} c^2_{12} c_{13} c_{23} s_{01} s_{03} s_{12} s^2_{13}
s_{23} \sin (\phi_x - 2\phi_y + \phi_z)
\nonumber \\
&& + c_{01} c_{02} c_{13} c_{23} s_{01} s^2_{02} s_{03} s^2_{12}
s^2_{13} s_{23} \sin (\phi_x - 2\phi_y + 3\phi_z) \; ,
\nonumber \\
J^{13}_{e\mu} & = & c_{02} c^2_{03} c_{12} c_{13} c_{23} s^2_{01}
s_{02} s_{03} s^2_{13} s_{23} \sin \phi_x +
c_{01} c_{02} c^2_{03} c_{12} c_{13} s_{01} s_{03} s_{13} s^2_{23}
\sin \phi_y
\nonumber \\
&& - c_{01} c^2_{03} c^2_{12} c^2_{13} c_{23} s_{01} s_{02} s_{13}
s_{23} \sin (\phi_x - \phi_y) + c_{01} c_{02} c^2_{03} c_{13} c_{23} s_{01}
s_{03} s_{12} s^2_{13} s_{23} \sin (\phi_x + \phi_z)
\nonumber \\
&& - c_{02} c^2_{03} c_{13} s^2_{01} s_{02} s_{03} s_{12} s_{13}
s^2_{23} \sin (\phi_y - \phi_z)
+ c_{01} c^2_{03} c^2_{13} c_{23} s_{01} s_{02} s^2_{12}
s_{13} s_{23} \sin (\phi_x - \phi_y + 2\phi_z)
\nonumber \\
&& - \left (c^2_{01} - s^2_{01} s^2_{02} \right )
c^2_{03} c_{12} c^2_{13} c_{23} s_{12} s_{13} s_{23}
\sin (\phi_x - \phi_y + \phi_z) \; ,
\nonumber \\
J^{23}_{e\mu} & = & - c_{02} c^2_{03} c_{12} c_{13} c_{23} s_{02}
s_{03} s^2_{13} s_{23} \sin \phi_x + c_{02} c^2_{03} c_{13} s_{02}
s_{03} s_{12} s_{13} s^2_{23} \sin (\phi_y - \phi_z)
\nonumber \\
&& + c^2_{02} c^2_{03} c_{12} c^2_{13} c_{23} s_{12} s_{13} s_{23}
\sin (\phi_x - \phi_y + \phi_z) \; ,
\end{eqnarray}
\normalsize
where
\begin{eqnarray}
\phi_x & \equiv & \delta_{03} - \delta_{02} - \delta_{23} \; ,
\nonumber \\
\phi_y & \equiv & \delta_{03} - \delta_{01} - \delta_{13} \; ,
\nonumber \\
\phi_z & \equiv & \delta_{02} - \delta_{01} - \delta_{12} \; .
\end{eqnarray}
With the help of Eq. (3), one may easily derive the
expressions of all the other rephasing invariants of $CP$ and $T$
violation from Eqs. (5), (6) and (7). The results obtained above
are new, and they are expected to be very useful for a systematic
study of $CP$- and $T$-violating effects in the four-neutrino mixing
models. The same results are also applicable for the discussion of
$CP$ and $T$ violation in the four-quark mixing models \cite{Chau}.

Note that all $CP$- and $T$-violating observables
in neutrino oscillations must be related linearly to $J^{ij}_{\alpha\beta}$.
To see this point more clearly, we consider that a neutrino $\nu_\alpha$
converts to another neutrino $\nu^{~}_\beta$ in vacuum.
The probability of this conversion is given by
\small
\begin{equation}
P(\nu_\alpha \rightarrow \nu^{~}_\beta) \; = \;
\delta_{\alpha\beta} - 4 \sum_{i<j} \left [ {\rm Re} \left (
V_{\alpha i} V_{\beta j} V^*_{\alpha j} V^*_{\beta i} \right )
\sin^2 F_{ji} \right ] - 2 \sum_{i<j} \left ( J^{ij}_{\alpha\beta}
\sin 2 F_{ji} \right ) \; ,
\end{equation}
\normalsize
where $F_{ji} \equiv 1.27 \Delta m^2_{ji} L/E$ with
$\Delta m^2_{ji} \equiv m^2_j - m^2_i$, $L$ stands for the baseline length
(in unit of km), and $E$ is the neutrino beam energy (in unit of GeV).
$CPT$ invariance assures that the transition probabilities
$P(\nu^{~}_\beta \rightarrow \nu_\alpha)$ and
$P(\overline{\nu}_\alpha \rightarrow \overline{\nu}^{~}_\beta)$ are
identical, and they can directly be read off from Eq. (9) through
the replacement $J^{ij}_{\alpha\beta} \Longrightarrow -J^{ij}_{\alpha\beta}$
(i.e., $V \Longrightarrow V^*$).
Thus the $CP$-violating asymmetries between
$P(\nu_\alpha \rightarrow \nu^{~}_\beta)$ and
$P(\overline{\nu}_\alpha \rightarrow \overline{\nu}^{~}_\beta)$
are equal to the $T$-violating asymmetries between
$P(\nu_\alpha \rightarrow \nu^{~}_\beta)$ and
$P(\nu^{~}_\beta \rightarrow \nu_\alpha)$. The latter can be explicitly
and compactly expressed as follows:
\begin{eqnarray}
\Delta P_{\alpha\beta} & \equiv &
P(\nu^{~}_\beta \rightarrow \nu^{~}_\alpha) ~ - ~
P(\nu^{~}_\alpha \rightarrow \nu^{~}_\beta) \;
\nonumber \\
& = & 16\left (J^{12}_{\alpha\beta} \sin F_{21} \sin F_{31} \sin F_{32}
+ J^{01}_{\alpha\beta} \sin F_{10} \sin F_{30} \sin F_{31}
\right .
\nonumber \\
&& ~ + \left . J^{02}_{\alpha\beta} \sin F_{20} \sin F_{30} \sin F_{32}
\right ) \; .
\end{eqnarray}
Equivalently, one may obtain
\begin{eqnarray}
\Delta P_{\alpha\beta} & = &
16\left (J^{23}_{\alpha\beta} \sin F_{21} \sin F_{31} \sin F_{32}
- J^{02}_{\alpha\beta} \sin F_{10} \sin F_{20} \sin F_{21}
\right .
\nonumber \\
&& ~ - \left . J^{03}_{\alpha\beta} \sin F_{10} \sin F_{30} \sin F_{31}
\right ) \; ,
\end{eqnarray}
or
\begin{eqnarray}
\Delta P_{\alpha\beta} & = &
16\left (J^{31}_{\alpha\beta} \sin F_{21} \sin F_{31} \sin F_{32}
+ J^{01}_{\alpha\beta} \sin F_{10} \sin F_{20} \sin F_{21}
\right .
\nonumber \\
&&  ~ - \left . J^{03}_{\alpha\beta} \sin F_{20} \sin F_{30} \sin F_{32}
\right ) \; .
\end{eqnarray}
In getting Eqs. (10) -- (12), the equality
$\sin 2 F_{ij} + \sin 2 F_{jk} + \sin 2 F_{ki}
= -4 \sin F_{ij} \sin F_{jk} \sin F_{ki}$ and Eq. (3) have been used.
Only three of the twelve asymmetries $\Delta P_{\alpha\beta}$
are independent, and they probe three of the six $CP$-violating phases
(or their combinations) of $V$. Since only the transitions between
active neutrinos can in practice be measured, we focus our interest on
three independent asymmetries of $CP$ and $T$ violation:
$\Delta P_{e \mu}$, $\Delta P_{\mu \tau}$ and $\Delta P_{\tau e}$.

The formulas of $\Delta P_{\alpha\beta}$ will remarkably be simplified,
if the hierarchy of neutrino mass-squared differences is taken into account.
For illustration, we assume that the current data of solar, atmospheric
and LSND neutrino oscillations can approximately be described by the
well-known (2+2) mixing scheme \cite{4N}. In this scheme the solar neutrino
problem is attributed essentially to the $\nu_e \rightarrow \nu_s$ oscillation
($\Delta m^2_{\rm sun} \approx \Delta m^2_{10}$),
the atmospheric neutrino anomaly arises dominantly from the
$\nu_\mu \rightarrow \nu_\tau$ oscillation
($\Delta m^2_{\rm atm} \approx \Delta m^2_{32}$), and the LSND neutrino
oscillation is governed by a bigger mass-squared difference
($\Delta m^2_{\rm LSND} \approx \Delta m^2_{21}$).
Without loss of generality, we have taken
$0 < m_0 < m_1 < m_2 < m_3$. The observed hierarchy
$\Delta m^2_{\rm sun} \ll \Delta m^2_{\rm atm} \ll \Delta m^2_{\rm LSND}$
allows us to make an analytical approximation for Eq. (11):
\begin{eqnarray}
\Delta P_{e \mu} & \approx & 16 \left ( J^{23}_{e \mu} \sin F_{\rm atm}
+ J^{01}_{e \mu} \sin F_{\rm sun} \right ) \sin^2 F_{\rm LSND} \; ,
\nonumber \\
\Delta P_{\mu \tau} & \approx & 16 \left ( J^{23}_{\mu \tau} \sin F_{\rm atm}
+ J^{01}_{\mu \tau} \sin F_{\rm sun} \right ) \sin^2 F_{\rm LSND} \; ,
\nonumber \\
\Delta P_{\tau e} & \approx & 16 \left ( J^{23}_{\tau e} \sin F_{\rm atm}
+ J^{01}_{\tau e} \sin F_{\rm sun} \right ) \sin^2 F_{\rm LSND} \; ,
\end{eqnarray}
where $(F_{\rm sun}, F_{\rm atm}, F_{\rm LSND}) =
1.27 (\Delta m^2_{\rm sun}, \Delta m^2_{\rm atm}, \Delta m^2_{\rm LSND}) L/E$.
If the magnitude of $J^{01}_{\alpha \beta}$ is comparable with or
smaller than that of $J^{23}_{\alpha \beta}$ (for $\alpha, \beta =
e, \mu, \tau$), then the asymmetries $\Delta P_{\alpha \beta}$ in Eq. (13)
are associated primarily with the oscillating term
$\sin F_{\rm atm} \sin^2 F_{\rm LSND}$.
If $|J^{01}_{\alpha \beta}| \gg |J^{23}_{\alpha \beta}|$, however,
the oscillation induced by $\sin F_{\rm sun} \sin^2 F_{\rm LSND}$ should
not be neglected in $\Delta P_{\alpha \beta}$.

To get a feeling of the relative magnitudes of
$J^{01}_{\alpha \beta}$ and $J^{23}_{\alpha \beta}$, we consider two
special but instructive cases for the mixing angles of $V$:

(a) $s_{02} \rightarrow 0$ and $s_{03} \rightarrow 0$. With the help
of Eqs. (5) -- (7), we arrive at
\begin{eqnarray}
J^{01}_{e\mu} & = & J^{01}_{\mu \tau} \; =\; J^{01}_{\tau e} \; =\; 0 \; ,
\nonumber \\
J^{23}_{e\mu} & = & J^{23}_{\mu \tau} \; =\; J^{23}_{\tau e} \; =\;
c_{12} c^2_{13} c_{23} s_{12} s_{13} s_{23} \sin \delta \; ,
\end{eqnarray}
where $\delta \equiv \phi_x - \phi_y + \phi_z = \delta_{13} - \delta_{12}
- \delta_{23}$. It turns out that the $T$-violating asymmetries in
Eq. (13) amount to one another and measure a common $CP$-violating
parameter, whose magnitude is identical to the well-known Jarlskog
invariant defined in the three-neutrino mixing scheme \cite{Jarlskog}.

(b) $s_{02}, s_{03}, s_{12}, s_{13} \sim \epsilon \ll 1$ \cite{Dai}.
In this more realistic case, we obtain
\begin{eqnarray}
J^{01}_{e \mu} & \approx &
c_{01} c_{23} s_{01} s_{03} s_{12} s_{23} \sin (\phi_x + \phi_z)
+ c_{01} c^2_{23} s_{01} s_{02} s_{12} \sin \phi_z
\nonumber \\
&& - c_{01} c_{23} s_{01} s_{02} s_{13} s_{23} \sin (\phi_x - \phi_y)
+ c_{01} s_{01} s_{03} s_{13} s^2_{23} \sin \phi_y \; ,
\nonumber \\
J^{01}_{\mu \tau} & \approx & 0 \; ,
\nonumber \\
J^{01}_{\tau e} & \approx &
c_{01} c_{23} s_{01} s_{03} s_{12} s_{23} \sin (\phi_x + \phi_z)
- c_{01} s_{01} s_{02} s_{12} s^2_{23} \sin \phi_z
\nonumber \\
&& - c_{01} c_{23} s_{01} s_{02} s_{13} s_{23} \sin (\phi_x - \phi_y)
- c_{01} c^2_{23} s_{01} s_{03} s_{13} \sin \phi_y \; ;
\end{eqnarray}
and
\begin{eqnarray}
J^{23}_{e \mu} & \approx & c_{23} s_{12} s_{13} s_{23} \sin\delta \; ,
\nonumber \\
J^{23}_{\mu \tau} & \approx & c_{23} s_{12} s_{13} s_{23} \sin\delta
+ c_{23} s_{02} s_{03} s_{23} \sin \phi_x \; ,
\nonumber \\
J^{23}_{\tau e} & \approx & c_{23} s_{12} s_{13} s_{23} \sin\delta \; ,
\end{eqnarray}
where the corrections of ${\cal O}(\epsilon^3)$ or smaller have been
neglected. Except $J^{01}_{\mu \tau} \sim 0$, the other five invariants
of $CP$ violation in Eqs. (15) and (16) are all suppressed by the factors
of ${\cal O}(\epsilon^2)$. If $\delta \sim {\cal O}(1)$ holds, then the
asymmetries $\Delta P_{\alpha \beta}$ in Eq. (13) are associated dominantly
with the oscillating term $\sin F_{\rm atm} \sin^2 F_{\rm LSND}$. An
interesting feature of $\Delta P_{\mu \tau}$ in case (b) is that it
depends primarily upon $J^{23}_{\mu \tau}$, whose magnitude gets
comparable contributions from the $\sin \delta$ and $\sin \phi_x$ terms.
Therefore these two $CP$-violating phases could in principle be
determined from the measurements of $\Delta P_{e \mu}$ and
$\Delta P_{\mu \tau}$.

In practice, however, the matter effects on neutrino mixing
parameters and neutrino oscillations must be taken into account.
To express the pattern of neutrino oscillations in matter in the
same form as that in vacuum, one may define the effective neutrino
masses $\tilde{m}_i$ ($i=0,1,2,3$) and the effective lepton flavor
mixing matrix $\tilde{V}$, in which the matter effects are already
included \cite{Xing01}. Then the matter-corrected
conversion probability of a neutrino $\nu_\alpha$ to another
neutrino $\nu_\beta$ can be written out in analogy to Eq. (9);
and the counterpart of the $T$-violating asymmetry
$\Delta P_{\alpha \beta}$ in matter is given, for instance, as
\begin{eqnarray}
\Delta \tilde{P}_{\alpha\beta} & = &
16\left (\tilde{J}^{23}_{\alpha\beta} \sin \tilde{F}_{21}
\sin \tilde{F}_{31} \sin \tilde{F}_{32} - \tilde{J}^{02}_{\alpha\beta}
\sin \tilde{F}_{10} \sin \tilde{F}_{20} \sin \tilde{F}_{21}
\right .
\nonumber \\
&& ~ - \left . \tilde{J}^{03}_{\alpha\beta} \sin \tilde{F}_{10}
\sin \tilde{F}_{30} \sin \tilde{F}_{31} \right ) \; ,
\end{eqnarray}
where $\tilde{F}_{ji} \equiv 1.27 \Delta \tilde{m}^2_{ji} L/E$,
$\Delta \tilde{m}^2_{ji} \equiv \tilde{m}^2_j - \tilde{m}^2_i$, and
$\tilde{J}^{ij}_{\alpha \beta} \equiv {\rm Im}
(\tilde{V}_{\alpha i} \tilde{V}_{\beta j} \tilde{V}^*_{\alpha j}
\tilde{V}^*_{\beta i})$. Note that $\Delta \tilde{m}^2_{ji}$ and
$\tilde{J}^{ij}_{\alpha \beta}$ depend upon the matter parameters
$a = \sqrt{2} G_{\rm F} N_e$ and $a' = \sqrt{2} G_{\rm F} N_n/2$,
where $N_e$ and $N_n$ denote the respective background densities
of electrons and neutrons \cite{MSW}. It seems very difficult, if
not impossible, to work out the analytically exact expressions
of $\Delta \tilde{m}^2_{ji}$ and $\tilde{J}^{ij}_{\alpha \beta}$
in terms of $a$ and $a'$ as well as the neutrino mixing parameters
in vacuum. Nevertheless,
a relationship between $J^{ij}_{\alpha \beta}$ and
$\tilde{J}^{ij}_{\alpha \beta}$ can be derived from the
equality between the commutator of lepton mass matrices in vacuum and
that in matter \cite{XingSUM}. The result is
\begin{eqnarray}
& & \Delta \tilde{m}^2_{10} \Delta \tilde{m}^2_{20}
\Delta \tilde{m}^2_{30}
\sum^3_{i=1} \left ( \tilde{J}^{0i}_{\alpha\beta} |\tilde{V}_{\gamma i}|^2 +
\tilde{J}^{0i}_{\beta\gamma} |\tilde{V}_{\alpha i}|^2 +
\tilde{J}^{0i}_{\gamma\alpha} |\tilde{V}_{\beta i}|^2
\right )
\nonumber \\
& & + \sum^3_{i=1} \sum^3_{j=1} \left [ \Delta \tilde{m}^2_{i0}
\left (\Delta \tilde{m}^2_{j0} \right )^2 \left (
\tilde{J}^{ij}_{\alpha\beta} |\tilde{V}_{\gamma j}|^2 +
\tilde{J}^{ij}_{\beta\gamma} |\tilde{V}_{\alpha j}|^2 +
\tilde{J}^{ij}_{\gamma\alpha} |\tilde{V}_{\beta j}|^2
\right ) \right ]
\nonumber \\
& = & \Delta m^2_{10} \Delta m^2_{20} \Delta m^2_{30} \sum^3_{i=1}
\left ( J^{0i}_{\alpha\beta} |V_{\gamma i}|^2 +
J^{0i}_{\beta\gamma} |V_{\alpha i}|^2 + J^{0i}_{\gamma\alpha} |V_{\beta i}|^2
\right )
\nonumber \\
& & + \sum^3_{i=1} \sum^3_{j=1} \left [ \Delta m^2_{i0}
\left ( \Delta m^2_{j0} \right )^2 \left (
J^{ij}_{\alpha\beta} |V_{\gamma j}|^2 +
J^{ij}_{\beta\gamma} |V_{\alpha j}|^2 + J^{ij}_{\gamma\alpha} |V_{\beta j}|^2
\right ) \right ] \; ,
\end{eqnarray}
where $(i,j,k)$ and $(\alpha, \beta, \gamma)$ run over
$(1,2,3)$ and $(e, \mu, \tau)$, respectively. Parametrizing
$\tilde{V}$ in terms of six effective mixing angles
$\tilde{\theta}_{ij}$ and six effective $CP$-violating phases
$\tilde{\delta}_{ij}$ (for $i, j =0,1,2,3$ and $i<j$) in analogy
to Eq. (4), one may obtain the explicit expressions of
$\tilde{J}^{ij}_{\alpha \beta}$ from Eqs. (5) -- (7) and rewrite
Eq. (18). If the mass of the sterile neutrino and its mixing with
active neutrinos are ``switched off'' (i.e., $a' = 0$, $m_0 =0$,
$\theta_{01} = \theta_{02} = \theta_{03} =0$, and
$\delta_{01} = \delta_{02} = \delta_{03} =0$), then the analytically
exact relations between the fundamental parameters in vacuum
($m_i$, $\theta_{12}$, $\theta_{13}$, $\theta_{23}$, $\delta$)
and their counterparts in matter
($\tilde{m}_i$, $\tilde{\theta}_{12}$, $\tilde{\theta}_{13}$,
$\tilde{\theta}_{23}$, $\tilde{\delta}$) can easily be
obtained \cite{Xing01}. In this case, Eq. (18) will be simplified
to an elegant form, the so-called Naumov identity \cite{Naumov}.

In summary, we have calculated the rephasing invariants of $CP$ and
$T$ violation in a favorable parametrization of the $4\times 4$
lepton flavor mixing matrix and derived their relations with the $CP$ and
$T$-violating asymmetries in neutrino oscillations. The matter
effects have been discussed to a limited extent. Our results are
expected to be useful for a systematic and model-independent analysis
of $CP$ and $T$ violation in the four-neutrino mixing scheme, in
particular, when sufficient data become available from the forthcoming
long-baseline neutrino oscillation experiments.

\vspace{0.3cm}

One of the authors (Z.Z.X.) is indebted to K. Hagiwara for warm
hospitality at KEK Theory Group, where part of this work was done.
He is also grateful to O. Yasuda for useful discussions during the
Tamura International School on Neutrino Physics.

\end{document}